\documentclass[12pt]{article}
\usepackage{epsfig}

\newcommand{\beq}{\begin{equation}}
\newcommand{\eeq}{\end{equation}}

\begin{document}
\title{Radiative decays of mesons in
the quark model: relativistic and
non-relativistic approaches}
\author{A.V. Anisovich, V.V. Anisovich and V.A. Nikonov}
\date{January 28, 2004}
\maketitle

\begin{abstract}
Different conclusions about quark nature of $f_0(980)$ based on the
analysis of data on the reaction
$\phi (1020)\to\gamma f_0(980)$ make it
necessary to perform detailed comparison of formulae used by
various groups for the calculation of partial widths of radiative
decays.  We carry out a comparative analysis of methods of calculation
of radiative decays like $\phi\to\gamma f_0$ and $\phi\to\gamma\eta$
performed by
F.E.~Close et~al., Phys. Rev. D {\bf65}, 092003 (2002), and
A.V.~Anisovich et~al., Eur. Phys. J. A {\bf12}, 103 (2001); Yad. Fiz.
{\bf65}, 523 (2002).
\end{abstract}

\section{Introduction}

At present there is a common understanding that radiative decays are
a powerful tool for the study of the quark structure of mesons,
and the calculation of
corresponding amplitudes is a subject of the increasing interest.
Several groups are involved in such
calculations, with different results and different conclusions
about the quark structure of mesons based on these
calculations. One of the main differences concerns the quark
structure of
$f_0(980)$ and $a_0(980)$. According to \cite{acha, fec}, $f_0(980)$
and $a_0(980)$ are dominantly $qq\bar q\bar q$ states, and the authors
claimed the
hypothesis about a dominant $q\bar q$ component
to be contradicting the data on
radiative decays. At the same time, in the papers of our group
\cite{epja,s,t}, an opposite conclusion is made:
the widths of radiative decays agree with the hypothesis about the
$q\bar q$ origin of $f_0(980)$ and $a_0(980)$, and the $s\bar s$
component in $f_0(980)$ should be about 60\%.  The sharp discrepancy of
conclusions demands to clarify the points, which resulted in
such a disagreement.

The recent paper \cite{kalash}, where the calculation of
a variety of reactions of one-photon radiative decays was carried out,
provided us with an opportunity to compare the methods of calculation.
Still, in \cite{kalash}, the calculation was performed in
non-relativistic approximation, while in \cite{epja,s,t}
we adopted a relativistic approach for the $q\bar
q$-systems. Let us point out that in \cite{s,t}, the formulae for the
amplitudes of radiative decays were written in a form allowing a
simple transformation to non-relativistic approximation for quark
composite systems. The result of comparison appeared to be startling: in
\cite{kalash}, the transition amplitudes depend explicitly on the
momentum of the decay particles, while in  \cite{s,t} there is no such
dependence. This circumstamce enforces us to perform a comparative
analysis of the two methods --- this is the subject of the present
paper.

We consider form factors of radiative decays involving
vector $(V)$, scalar  $(S)$ and pseudoscalar $(P)$ mesons, such as
$V\to\gamma P$ and $V\to\gamma S$, written in terms of Feynman
integrals; then we transform them to non-relativistic approxomation.
In the framework of such a procedure, one can see directly that in the
amplitude of radiative decays $V\to\gamma P$ and
$V\to\gamma S$ the momentum depependence, like that obtained in
\cite{kalash}, is absent. The momentum dependence of the matrix
element implemented in \cite{kalash} causes the violation of
Lorentz-invariance: the requirement of Lorentz-invariance, due to
zero photon mass, reveals itself also in a non-relativistic description
of composite systems.

Let us clarify this statement in more detail. The Lorentz-invariant
structure of the amplitudes of the transitions $V\to\gamma P$
and $V\to\gamma S$ have the form:
\begin{eqnarray}
\label{m1}
A_{V\to\gamma P} &=& -i
\varepsilon_{\mu\alpha\nu_1\nu_2}
\epsilon^{(V)}_\mu \epsilon^{(\gamma)}_\alpha
q_{\nu_1}p_{\nu_2}F_{V\to\gamma P}(q^2) \ ,
\nonumber \\
A_{V\to\gamma S} &=&
\left ( g_{\mu\alpha} - \frac{q_\mu p_\alpha}{(pq)}\
\right )
\epsilon^{(V)}_\mu \epsilon^{(\gamma)}_\alpha
\,F_{V\to\gamma S}(q^2)\ .
\label{m2}
\end{eqnarray}
Here $\varepsilon_{\mu\alpha\nu_1\nu_2}$ is the antisymmetrical tensor,
$\epsilon^{(V)}$ and $\epsilon^{(\gamma)}$ are polarization
four-vectors for the vector meson and photon, and
$p$ and $q$ are their four-momenta;
$g_{\mu\alpha}$ is the metric tensor.
Invariant form factor
$F_{V\to\gamma P}(q^2)$ (or  $F_{V\to\gamma S}(q^2)$) is a
function of three independent variables: $m^2_V$, $q^2$ and $m^2_P$
(or $m^2_S$).  For real decays $q^2=0$, and this circumstance
simplifies calculations of the form factors $F_{V\to\gamma P}(0)$
and $F_{V\to\gamma S}(0)$ as convolutions of quark wave functions.
Form factors represented in terms of
 light-cone variables $k_\perp$ and
$x$ read as follows \cite{s,t}:
\begin{eqnarray}
&& F_{V\to\gamma
P}(0)\ \sim\ \psi_V(k_\perp,x)\ \otimes\ \psi_P(k_\perp,x)\ ,
\label{m3}
\nonumber \\
&& F_{V\to\gamma S}(0)\ \sim\ \psi_V(k_\perp,x)\ \otimes\
\psi_S(k_\perp,x)\ ,
\label{m4}
\end{eqnarray}
and in these convolutions the photon momentum  $q$ is not
present explicitly. In the relativistic form factor
representation, such a property is obvious because of the Lorentz
invariance:  form factors with  $q^2=0$ allow one to choose any set
of variables,  $q_0=|{\bf q}|=0$ included.

After coming to non-relativistic approach, the form factors obtained in
\cite{epja,s,t} turn into the convolutions of quark wave functions:
\beq \label{m5}
\psi_1(|{\bf k}|)\ \otimes\ \psi_2(|{\bf k}|)\ ,
\eeq
where $\bf k$ is relative momentum of the $q\bar q$-system. Another
representation of form factors is declared in
\cite{kalash}: form factors for the transitions
$V\to\gamma P$ and $V\to\gamma S$ are written as convolutions
\beq
\label{m6}
\psi_1(|{\bf k}|)\ \otimes\ \psi_2\left(\left|{\bf
k}+\frac12 {\bf q} \right|\right) \ ,
\eeq
(to avoid misunderstanding, let us note that in \cite{kalash} the
photon momentum was denoted as $\bf p$, see Eq. (2) of this paper, but
here we use ${\bf q}$ for it).

The formulae (\ref{m5}) and (\ref{m6}) reflect the discrepancy
between papers \cite{epja,s,t} and \cite{kalash}, so they are the
subject of the present discussion.

By discussing non-relativistic formulae for the transitions
$V\to\gamma P$ and $V\to\gamma S$, it would be more transparent
not to use the trick with the choice of variables but to
work directly with physical momenta
$q_0=|{\bf q}|\neq0$.  Such a calculation is done in the next
Section by studying form factors  $F_{V\to\gamma P}(0)$ and
$F_{V\to\gamma S}(0)$ in the non-relativistic limit. We show that, by
making a correct transformation, when in the integrand the terms of
different order are not kept, the radiative decay form factors are
defined by the convolution   (\ref{m5}).

Another difference between papers \cite{kalash}  and
\cite{epja,s,t}
consists in a way of the
application of the threshold theorem \cite{siegert}
($F_{V\to\gamma S}(0)\sim\omega$ at small mass difference of the vector
and scalar particles, $\omega=m_V-m_S$ ) to the decay
$\phi(1020)\to\gamma f_0(980)$. In \cite{epja,s,t}
the process $V\to\gamma S$ was considered in terms of the
additive quark model when the photon is emitted by one of the
constituent quarks. For this process
the equality $F_{V\to\gamma S}(0)=0$ is impossible if $V$ and $S$
are basic states ($1^3S_1q\bar q$ and $1^3P_0q\bar q$), for
radial wave functions of these states have no nodes and the convolution
(\ref{m5}) does not change sign. For basic states
the requirement $F_{V\to\gamma S}(0)=0$
is fulfilled only under the inclusion of  processes
beyond the additive model \cite{maxim}, for example, it can be
the photon emission by the $t$-channel exchange current \cite{tt}.
In \cite{kalash}
the threshold  behaviour,
$F_{V\to\gamma S}(0)\sim\omega$ at $\omega\to0$ (see Eq.
(10) of paper \cite{kalash}), is believed to be
due to the proper choice of the wave  functions.

We do not investigate here the threshold behaviour of form factors
in detail --
special articles are devoted to this subject \cite{maxim,tt}, where
two types of reactions have been considered: with the production
of a bound system and with production of two unbound particles with
the same quantum numbers, for example,
$\phi(1020)\to\gamma f_0(980)$
and  $e^+e^-\to\phi\to\gamma(\pi\pi)_S$.
The amplitudes of these reactions are
connected with each other, namely, the transition amplitude
of  $\phi(1020)\to\gamma f_0(980)$ is
a residue in the poles of the amplitude of the process
$e^+e^-\to\phi\to\gamma(\pi\pi)_S$:
\beq \label{m7}
A_{e^+e^-\to\phi\to\gamma(\pi\pi)_S}(s_{e^+e^-}\ ,s_{\pi\pi},0)\ =\
\frac{A_{\phi(1020)\to\gamma f_0(980)}
(0)}{(m^2_\phi-s_{e^+e^-})(m^2_{f_0}-s_{\pi\pi})}
+\mbox{ smoother contributions ,}
\eeq
where $s_{e^+e^-}$ and $s_{\pi\pi}$ are the effective masses
squared of the
$e^+e^-$ and $\pi\pi$ systems.
The threshold theorem is valid for the reactions with stable
particles, that is for
$A_{e^+e^-\to\phi\to\gamma(\pi\pi)_S}(s_{e^+e^-}\ ,s_{\pi\pi},0)$
and $A_{\phi(1020)\to\gamma (\pi\pi)_S}(m^2_\phi,s_{\pi\pi},0)$:
\beq \label{m8}
A_{e^+e^-\to\phi\to\gamma(\pi\pi)_S}(s_{e^+e^-}\ ,s_{\pi\pi},0)\ \sim\
(s_{e^+e^-}-s_{\pi\pi})\ ,
\eeq
$$
A_{\phi(1020)\to\gamma (\pi\pi)_S}(m^2_\phi,s_{\pi\pi},0) \sim \
(m^2_\phi-s_{\pi\pi}),
$$
for the $\phi(1020)$ can be considered as a stable particle with a good
accuracy.

The problem of direct application of the threshold theorem to the
$f_0(980)$ is not self-evident because this resonance definitely cannot
be considered as a stable particle: it is characterized by two poles
on different sheets related to the $K\bar K$ threshold, at $1020-i40$
MeV and $960-i200$ MeV, and the second pole is essential for
the formation of the $f_0(980)$ signal, see \cite{tt} for more detail.

\section{Form factors $F_{V\to\gamma S}(0)$ and $F_{V\to\gamma P}(0)$
in additive quark model}

In the additive quark model, radiative decay is a three-stage process:
the transition $V\to q\bar q$, photon emission by one of the
quarks and the fusion of quarks into a final meson
($S$ or $P$), see Fig.1. The processes considered here,
$V\to\gamma S$ and $V\to\gamma P$,  are the transitions of both
electric and magnetic types. Accordingly,
the formula for the photon--quark vertex reads:
\beq \label{m9}
\frac1{2m}\,(k_{1\alpha}+k'_{1\alpha})+\sigma_{\alpha\beta}\,
\frac{q_\beta}{2m}\ ,
\eeq
where $m$ is the quark mass,
$\sigma_{\alpha\beta}
=(\gamma_\alpha\gamma_\beta -\gamma_\beta\gamma_\alpha)/2$,
for the notations of momenta see Fig. 1.
Such a representation of the vertex
being eqivalent to the expression with the use of  $\gamma_\alpha$
simplifies the calculations related to the
transform to non-relativistic limit.

In the calculations, we work with amplitudues written as
$\epsilon^{(\gamma)}_\alpha\epsilon^{(V)}_\mu
A_{\mu\alpha}^{V\to\gamma \, P/S}$ taking into account
the requirements
$\epsilon^{(\gamma)}_\alpha q_\alpha =0$ and $\epsilon^{(V)}_\mu
p_\mu =0$. Therefore, the calculated amplitudes obey the constraints
$q_\alpha A_{\mu\alpha}^{V\to\gamma\, P/S}=0$ and
$p_\mu A_{\mu\alpha}^{V\to\gamma \, P/S}=0$.

\subsection{Feynman representation of the triangle diagram of Fig. 1}

Feynman integral for the diagram of Fig. 1 reads:
\beq \label{m10}
\epsilon^{(\gamma)}_\alpha\epsilon^{(V)}_\mu\int \frac{d^4k}{i(2\pi)^4}\, G_V
\frac{(-){\rm Sp}\left[\gamma_\mu (\hat k_1+m)\Gamma_\alpha(\hat
k'_1+m)\Gamma  (-\hat k_2+m)\right]}{(m^2-k^2_1-i0)(m^2-k'^2_1-i0)(m^2
-k^2_2-i0)}\, G\ ,
\eeq
where for the transition $V\to\gamma S$ the spin operators
$\Gamma$ and $\Gamma_\alpha$ are defined as follows:
\beq \label{m11}
\Gamma=I\ , \qquad \Gamma_\alpha=\frac1{2m}(k_{1\alpha}+k'_{1\alpha})\ ,
\eeq
and for $V\to\gamma P$ they are equal to
\beq \label{m12}
\Gamma\ =\ \gamma_5\ , \qquad \Gamma_\alpha=\ \sigma_{\alpha\beta}
\frac{q_\beta}{2m}\ .
\eeq
Transition vertices $V\to q\bar q$ and $q\bar q\to S$ (or $q\bar q\to
P$) are denoted as $G_V$ and $G$.

A suitable tranformation procedure for getting
 non-relativistic expression
is to introduce in (\ref{m10}) the two-component spinors for quark and
antiquark,
$\varphi_j$ and $\chi_j$, that is realized by susbstituting
\beq \label{m13}
(\hat k_1+m)\to\sum_{j=1,2} u^j(k_1)\bar u^j(k_1)\ , \quad (\hat
k'_1+m)\to\sum_{j=1,2} u^j(k'_1)\bar u^j(k'_1)\ ,
\eeq
where
\beq \label{m14}
u^j(k)\ =\ \sqrt{k_0+m} \left({\varphi_j \atop
\frac{\mbox{\boldmath$\sigma k$}}{k_0+m}\ \varphi_j} \right),\ j=1,2,
\eeq
and
\beq \label{m15}
\left(\hat k_2-m\right)\ \rightarrow\ \sum_{j=3,4} u^j(-k_2)\bar
u^j(-k_2)\ ,
\eeq
where
\beq \label{m16}
u^j(-k)\ =\ i\sqrt{k_0+m}\left({ \frac{\mbox{\boldmath$\sigma k$}}{
k_0+m}\ \chi^j \atop \chi^j } \right).
\eeq
After substitutions (\ref{m13})--(\ref{m16}), we have
the two-dimensional trace in the numerator of the integrand
(\ref{m10}).

\subsection{The diagram of Fig.1 in non-relativistic approximation}

Now we go to non-relativistic approximation in the vector-particle rest
frame. Denoting the four-momentum of vector particle as
$p=(p_0,{\bf p}_\perp,p_z)$, we have in this frame:
\beq \label{m17}
p\ =\ (2m-\varepsilon_V,0,0)\ ,
\eeq
where $\varepsilon_V$ is the binding energy of vector particle, which
is supposed to be small as compared to the quark mass,
$\varepsilon_V\ll m$.

Let the photon fly along the  $z$-axis, then
\beq \label{m18}
q\ =\ (q_z,0,q_z)\ ,
\eeq
and the polarization vector of photon lays in the $(x,y)$-plane.

The four-momentum of scalar (pseudoscalar) particle is equal to
\beq \label{m19}
p'=(2m-\varepsilon_V-q_z,0,-q_z)=\left(\sqrt{(2m-\varepsilon)^2+q^2_z}
,0,-q_z\right)\simeq\left(2m-\varepsilon+\frac{q^2_z}{2m},0,-q_z\right).
\eeq
Here $\varepsilon$ is the binding energy of scalar (pseudoscalar)
particle, which is also small compared to the mass of constituent
$\varepsilon\ll m$ .

\subsection{The reaction $V\to\gamma S$}

The transition to non-relativistic approximation in the numerator
of the integrand (\ref{m10}) provides the following formula for the
reaction
$V\to\gamma S$:
\beq
\label{m20}
-\mbox{ Sp}_2\left[2m\sigma_\mu\cdot({\bf k}_2-{\bf k}_1')
\mbox{\boldmath$\sigma$}\right](k_{1\alpha}+k'_{1\alpha})\ .
\eeq
The notation Sp$_2$ stands for the trace of
two-dimensional matrices. In (\ref{m20}),
in the transition to non-relativism, the following terms are of the
leading order.\\
In the photon--quark vertex (electric interaction):
\beq
\label{m21}
\bar u(k_1)\,\frac{k_{1\alpha}+k'_{1\alpha}}{2m}\ u(k'_1)\
\longrightarrow\
\varphi^+(1)(k_{1\alpha}+k'_{1\alpha})\varphi(1')\ ,
\eeq
in the vertex
$V\to q\bar q$:
\beq
\label{m22}
\bar u(-k_2)\gamma_\mu\,u(k_1)\
\longrightarrow\ \chi^+(2)\, 2m\sigma_\mu\,\varphi(1)\ ,
\eeq in the vertex
$q\bar q\to S$:
\beq
\label{m23}
\bar u(k'_1)\,u(-k_2)\
\longrightarrow\ \varphi^+(1') \mbox{\boldmath$\sigma$} ({\bf k}_2-{\bf
k}'_1)\,\chi(2)\ .
\eeq
The constituent propagators in the transition to non-relativism
should be replaced as follows:
\beq
\label{m24}
(m^2-k^2-i0)^{-1}\ \longrightarrow\ (-2mE+{\bf k}^2-i0)^{-1}\ ,
\eeq
where $E=k_0-m$ and $m^2-k^2_0\simeq-2mE$. Then the amplitude for
the transition $V\to\gamma S$ is defined
by the diagram of Fig. 1, it reads as follows:
\beq
\label{m25}
\epsilon^{(V)}_\mu\epsilon^{(\gamma)}_\alpha
\int \frac{dEd^3k}{i(2\pi)^4}\,G_V\, \frac{-{\rm
Sp}_2[2m\sigma_\mu\cdot({\bf k}_2-{\bf k'}_1) \mbox{\boldmath$\sigma$}
](k_{1\alpha}+k'_{1\alpha})}{(-2mE_1 + {\bf k}_1^2-i0)(-2mE'_1+{\bf
k}'^2_1-i0)(-2mE_2+{\bf k}_2^2-i0)}\, G\ .
\eeq
Furthermore, we  denote
$E_2\equiv E$, ${\bf k}_2\equiv{\bf k}$. With these notations
the energy--momentum consevation laws
are written as follows:
\begin{eqnarray}
 E_1=-\varepsilon_V-E\ , \quad && {\bf k}_1=-{\bf k}\ , \nonumber\\
 E'_1=-\varepsilon_V-E-q_z\ , \quad && {\bf k}'_1=-{\bf k}-{\bf q}\ .
\label{m26}
\end{eqnarray}
One can integrate over $E$ by closing the integration contour in the lower
half-plane of the complex-valued $E$, that is equivalent to
substitution in (\ref{m25}):
\beq \label{m27}
(-2mE+{\bf k}^2-i0)^{-1}\ \longrightarrow\ \frac{2\pi i}{2m}\
\delta\left(E-\frac{{\bf k}^2}{2m}\right) .
\eeq
By fixing $E={\bf k}^2/2m$,
we can evaluate the order of value of momenta
entering (\ref{m25}). Formula ({\ref{m19}) gives us
$2m-\varepsilon_V-q_z=2m-\varepsilon+q^2_z/2m^2$, or
\beq \label{m28}
q_z\ \simeq \ \varepsilon_V-\varepsilon\ ,
\eeq
because $q^2_z/2m$ is the magnitude of the next-to-leading
order of value. In the integrand of (\ref{m25}),
the momentum square is of the order of
\beq \label{m29}
{\bf k}^2\ \sim\ 2m\varepsilon\ \sim\ 2m\varepsilon_V\ ,
\eeq
and this means that
\beq \label{m30}
q_z\ \ll\ |{\bf k}|\ .
\eeq
Therefore, within non-relativistic approximation, the amplitude for the
transition $V\to\gamma S$ reads:
\beq \label{m31}
\epsilon^{(V)}_\mu \epsilon^{(\gamma)}_\alpha \int \frac{d^3k}{(2\pi)^3}\, \psi_V
(k)\psi_S(k)\,\frac1{2m}(-4)\mbox{ Sp}_2[2m\sigma_\mu\cdot 2(
\mbox{\boldmath$k\sigma$})](-2k_\alpha)\ ,
\eeq
where the requirement (\ref{m30}) is duely taken into account and the
wave functions for vector and scalar particle are introduced:
\beq \label{m32}
\psi_V(k)\ =\ \frac{G_V}{4(m\varepsilon_V+{\bf k}^2)}\ , \quad
\psi_S(k)\ =\ \frac G{4(m\varepsilon+{\bf k}^2)}\ .
\eeq
Recall, polarization vector $\epsilon^{(V)}$
do not contain the time-like
component and polarization vector of the photon belongs to the
$(x,y)$-plane.
Accounting for Sp$_2[\sigma_\mu\sigma_\beta]=2\delta_{\mu\beta}$, where
$\delta_{\mu\beta}$ is the three-dimensional Kronecker symbol, and
substituting in the integrand
 $k_\mu k_\alpha\to\delta_{\mu\alpha}{\bf k}^2/3$,
we have had final expression:
\beq \label{m33}
A_{V\to\gamma S}=\left(\epsilon^{(V)}\epsilon^{(\gamma)}\right)\int
\frac{d^3k}{(2\pi)^3}\,\psi_V(k)\,\psi_S(k)\,\frac{32}3\,k^2
=\left(\epsilon^{(V)}\epsilon^{(\gamma)}\right)
\int\limits^\infty_0
\frac{dk^2}\pi\,\psi_V(k)\psi_S(k)\frac8{3\pi}\,k^3\ .
\eeq
Here we re-denoted ${\bf k}^2 \to k^2$.
The photon polarization belongs to the plane orthogonal
to $p$ and $q$, so one can re-write (\ref{m33}) in the form
of Eq. (\ref{m1}) by using
$$
\label{31a}
\left ( g_{\mu\alpha} -  \frac{q_\mu p_\alpha}{(pq)}
\right )
\epsilon^{(V)}_\mu \epsilon^{(\gamma)}_\alpha =
(\epsilon^{(V)} \epsilon^{(\gamma)}).
\nonumber
$$
Formula (\ref{m33})
coincides with that obtained in terms of the double
spectral-integral representation (Eq. (32) of \cite{s}), if in the
relativistic-invariant formula
(32) of \cite{s} one makes a transform to
the non-relativistic limit.

\subsection{The reaction $V\to\gamma P$}

In the reaction $V\to\gamma P$, non-relativistic spin
factor (numerator of the integrand of (\ref{m10}) has the form:
\beq \label{m34}
(-)\mbox{ Sp}_2[2m\sigma_\mu\cdot i\varepsilon_{\alpha\beta\gamma} \
q_\beta \sigma_\gamma\cdot2m]\ =\ -i8m^2 \varepsilon
_{\mu\alpha\beta}\  q_\beta\ ,
\eeq
where $\varepsilon_{\alpha\beta\gamma}$ is three-dimensional
antisymmetric tensor.
As a result, we have:
\beq
\label{m35}
A_{V\to\gamma P} = -i
\varepsilon_{\mu\alpha\nu_1\nu_2}
\epsilon^{(V)}_\mu \epsilon^{(\gamma)}_\alpha
q_{\nu_1}p_{\nu_2}F_{V\to\gamma P}(q^2)
=-i\varepsilon
_{\mu\alpha\beta}\epsilon^{(V)}_\mu\epsilon^{(\gamma)}_\alpha q_\beta
\int\limits^\infty_0 \frac{dk^2}\pi\,\psi_V(k)\psi_P(k)
\frac{4km}\pi\ .
\eeq
This expression coincides also with its relativistic-invariant counterpart
obtained in \cite{s} (Eq. (61)), if in \cite{s} one makes
a transition to non-relativistic limit.

\subsection{Normalization of wave functions  $\psi_V(k)$, $\psi_S(k)$
and $\psi_P(k)$}

Here we use  the wave function normalization
similar to what was applied in  \cite{epja,s,t} for the
relativistic treatment of composite systems.

For scalar particle, the normalization reads:
\beq \label{m36}
\int\limits^\infty_0\ \frac{dk^2}\pi\ \psi^2_S(k)\frac{2k^3}{\pi m}\ =\
1\ ,
\eeq
and for vector particle it is
\beq \label{m37}
\int\limits^\infty_0\ \frac{dk^2}\pi\ \psi^2_V(k)\frac{2km}\pi\ =\ 1\ .
\eeq
Normalization for pseudoscalr particle is the same as for vector one: in
(\ref{m37}) one should replace $\psi_V(k)\to\psi_P(k)$.

Normalization condition of wave function can be re-formulated
as requirement for charge form factor at $q^2=0$:
\beq
\label{m38}
F_{charge}(0)\ =\ 1\ .
\eeq
It is easy to see that formulae  (\ref{m36})  and
(\ref{m37}) are actually the requirements for charge form factors. Let
us explain this using scalar particle as an example.

Form factor of scalar particle is defined by triangle diagram of
Fig. 1 type. Using the same calculation technique, which resulted in
formula (\ref{m31}), we obtain:
\beq
\label{m39}
F^{(S)}_{charge}(0)\ =\int \frac{d^3k}{(2\pi)^3}\, \psi^2_S(k)\
\frac2m\mbox{ Sp}_2[2(\mbox{\boldmath$k\sigma$})\cdot 2
(\mbox{\boldmath$k\sigma$})]\frac12\ \quad .
\eeq
Here, as by derivation of Eq. (\ref{m31}), the factor $2/m$ arises
due to the integration over $E$ and wave function definition
$\psi_S(k)$; the vertex $S\to q\bar q$ is equal to
$2(\mbox{\boldmath$k\sigma$})$, and the factor 1/2 appeared
because of substitution
$k_{1\alpha}+k'_{1\alpha} \to (p_{1\alpha}+p'_{1\alpha})/2$: at
$\alpha=0$, that corresponds to the interaction with Coulomb field,
 we have $k_{10}=k'_{10}\simeq m$ and
$p_{10}=p'_{10}\simeq2m)$.  Just the condition
$F^{(S)}_{charge}(0)=1$ gives us the formula (\ref{m36}).

\section{Comments on the analyticity of the amplitude $V \to \gamma S $}

{\bf 1.}
The threshold theorem can be reformulated as the requirement of
analyticity of the field theory amplitude. The transition amplitude
$V\to \gamma S$ written for elementary particles of the field theory
Lagrangian has a structure as follows:
\beq
 \label{2}
 \left
(g_{\mu\alpha}-\frac{2 q_\mu p_\alpha}{s_V-s_S}\right ) A(s_V,s_S,0)\ ,
\eeq
where $(pq) =(s_V-s_S)/2$ and $q^2=0$.
Then analyticity constraint reads
\beq                    \label{3}
\left [ A(s_V,s_S,0)\right ]_{s_V\to s_S} \to 0\ .
\eeq
The constraint (\ref{3}) is written for the particles which form the
standard sets of $|in \rangle $ and $\langle out |$ states, i.e. for
particles which can be considered as stable ones. The problem of
applying (\ref{3}) to resonances is questionable because the resonance
amplitude is determined by the pole residue, while the analytisity
requirement (\ref{m7}) includes smoother terms as well.

{\bf 2.}
The amplitude of additive quark model for the transition
$V\to \gamma S$ determined by (\ref{m33}) cannot be equal to zero.
Indeed, to turn the right-hand side of Eq.
(31) to zero one would need at least that one wave function was a
sign-changing one. But wave functions of the lowest radial-excitation
states, with radial quantum number $n=1$, have no nodes. Therefore, the
transition matrix element  for basic states with $n=1$ does not turn
to zero, this fact not depending on what potential for $V$ and
$S$ systems is under consideration.

Although in the additive approach
$A_{V\to \gamma S}(0)\neq 0$, the factor  $(m_V-m_S)$
inherent in the E1-transition can be extracted explicitly from the
right-hand side of (\ref{m33}) for a wide class of models.

In order to clarify this point let us consider as an example the
exponetial approximation for the wave functions $\psi_V(k^2)$ and
$\psi_S(k^2)$:
\beq
\psi_V(q)\sim\exp\left(-b_Vk^2\right), \quad
\psi_S(q)\sim\exp\left(-b_Sk^2\right).
\label{11a}
\eeq
With exponential wave functions, the matrix element for the
$V\to\gamma S$ amplitude
given by the additive quark model diagram, Eq. (\ref{m33}), up to
a numerical factor is equal to:
\beq
\label{11c}
A_{V\to\gamma S}\sim
(\epsilon^{(V)}\epsilon^{(\gamma)})\,
 \frac{b^{3/4}_Vb^{5/4}_S}{(b_V+b_S)^{5/2}}\ .
\eeq
In case of
one-flavour quarks in the exponetial potential well one has
\beq
\label{11e}
m(m_V-m_S)\ =\ b^{-1}_V\ ,
\eeq
that  gives the amplitude
$A_{V\to\gamma S}$ in the dipole
emission  representation:
\beq
\label{11d}
A_{V\to\gamma S}\sim
(\epsilon^{(V)}\epsilon^{(\gamma)})
\frac{b^{7/4}_Vb^{5/4}_S}{(b_V+b_S)^{5/2}}
\, m(m_V-m_S)\ .
\eeq
In case under consideration
the factor ($m_V-m_S)$, or ($\varepsilon_S-\varepsilon_V)$,
relates to the difference between the
$V$ and $S$ levels and is defined by the $b_V$ value only.
The amplitude $A_{V\to \gamma S}(0)$ turns to zero when
binding energies of either  $V$-meson or $S$-meson tend to
zero: $\varepsilon_V \to 0$ or $\varepsilon_S \to 0$.

The considered example demonstrates that the statement of Ref.
\cite{acha} that the triangle diagram contribution does not contain
the factor ($\varepsilon_S-\varepsilon_V)$ is not generally justified.
Still, the equivalence of the additive model and dipole representations
for transition amplitudes depends on the type of interaction under
consideration, see \cite{tt} for more detail.

{\bf 3.} There is one more point that is worth mentioning in connection
with the present discussion, that is a representation of the spin
operator. In \cite{maxim}, it was emphasized that the choice of the
spin operator (\ref{2}) is not  unique as sometimes was it suggested
\cite{acha}. For the spin structure of the amplitude $V\to \gamma S$
with emission of a real photon ($q^2=0$),
one can alternatively use the metric tensor
$g^{\perp\perp}_{\mu\alpha}(0)$
defined in the space orthogonal to the four-momenta $p$ and $q$:
\beq
\label{4}
g^{\perp\perp}_{\mu\alpha}(0)\ =\
g_{\mu\alpha }+\frac{4\,s_V}{(s_V-s_S)^2}\,q_{\mu}q_\alpha
-\frac{2}{s_V-s_S}\,(p_{\mu}q_\alpha +q_{\mu} p_\alpha)\ .
\eeq
The matter
is that the difference
\beq
\label{5}
g_{\mu\alpha}^{\perp\perp}(0)-
\left (g_{\mu\alpha}-\frac{2 q_\mu p_\alpha}{s_V-s_S}\right )
=4L_{\mu\alpha}(0)\ ,
\eeq
where
\beq    \label{6}
L_{\mu\alpha}(0)\ =\
\frac{s_V}{(s_V-s_S)^2}\,q_{\mu}q_\alpha
-\frac{1}{2(s_V-s_S)}\,p_{\mu}q_\alpha\ ,
\eeq
is a nilpotent operator:
\beq
\label{9}
L_{\mu\alpha}(0)\, L_{\mu\alpha}(0)\ =\ 0\   .
\eeq
Adding a nilpotent operator
to a spin operator one does not change the definition of the
transverse amplitudes, see \cite{maxim} for detal.

{\bf 4.} One can rise the question whether it is possible
to re-define
the radiative decay form factors
(for $V\to \gamma S$ and other similar processes) in accordance
with Eq.  (\ref{m6}). For the decay $V\to \gamma S$,
we have $|{\bf q }|= m_V-m_S$, so the re-definition does not affect the
fact that
$F_{V\to\gamma S}(m^2_V,m^2_S, q^2=0)$ depends on $m^2_V$ and $m^2_S$
only \cite{melikhov}.
However, it is necessary to take into consideration that the
convolution (\ref{m6}) represents also a standard  determination
of the transition form factor for the space-like momenta
$q^2= -{\bf q}^2 < 0$. So, within this re-definition and
applying $q^2=-|{\bf q }|^2= -(m_V-m_S)^2$ one has the equality
$$
F_{V\to\gamma S}(m^2_V,m^2_S, 0)=
F_{V\to\gamma S}(m^2_V,m^2_S, -(m_V-m_S)^2)\ ,
$$
which does not look
reliable.

\section{Conclusion}

We have obtained formulae, within a non-relativistic quark
model, for the transitions
$V\to\gamma S$ and $V\to\gamma P$ starting from Feynman integrals. We
have shown that, after a correct transition to the non-relativistic
approximation, the decay amplitude with the emission of a real photon
 $(q^2=0)$ is determined by the
convolution of wave functions (see (\ref{m33}) and
(\ref{m35})), where there is no photon-momentum dependence.
This is a natural consequence of the Lorentz-invariant structure of
the transition amplitudes, and it is a common property
independent of whether we use relativistic or non-relativistic
representations of the amplitude, see the discussion of formulae
(3) and (4) here, as well as Eqs. (32) and (61) of paper \cite{s}.
In this point our formulae differ from those of the paper
\cite{kalash}, where there is a  $\bf q$-dependence in the quark
wave functions.

Another difference between papers \cite{epja,s,t} and
\cite{acha,fec,kalash} consists in the application of the threshold
theorem \cite{siegert} (see also \cite {8,9}) to the decay
$\phi(1020)\to\gamma f_0(980)$ -- we discuss this problem
in separate publication \cite{tt}.

We thank
Ya.I. Azimov, L.G. Dakhno, S.S. Gershtein, Yu.S. Kalashnikova,
A.K. Likhoded, M.A. Matveev,
V.A. Markov, D.I.  Melikhov, A.V. Sarantsev and W.B. von Schlippe for
helpful and stimulating discussions of problems involved.  This work is
supported by the RFBR Grant N 0102-17861.


\begin{figure}[h]
\centerline{\epsfig{file=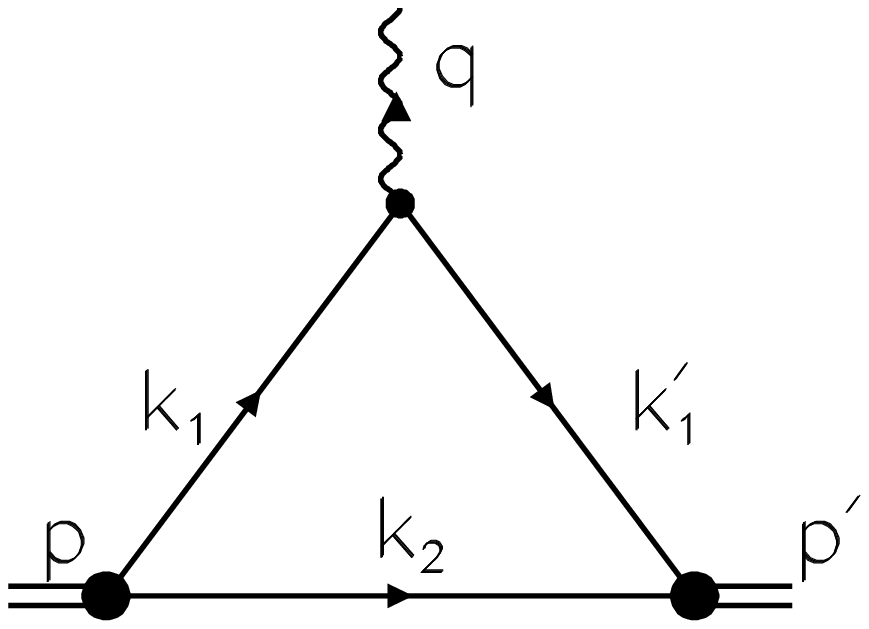,width=12cm}}
\caption{Quark diagram for the transition form factor $V\to\gamma S/P$.}
\end{figure}

\end{document}